\documentclass{emulateapj}

\def\simg{\mathrel{\rlap{\raise 0.511ex \hbox{$>$}}{\lower 0.511ex \hbox{$\sim$}}}}
\def\siml{\mathrel{\rlap{\raise 0.511ex \hbox{$<$}}{\lower 0.511ex \hbox{$\sim$}}}}
\def\eps{\varepsilon} \def\LAE{Large-Angle Emission } \def\tg{t_\gamma} \def\deg{^{\rm o}}
\def\ds{\displaystyle} \def\inu{i'_{\nu'}}
\def\h{\hspace*{-1mm}} \def\hh{\hspace*{-2mm}}

\begin{document}

\parskip 5pt

\title{X-ray Afterglows from the Gamma-Ray Burst "Large(r)-Angle" Emission }

\author{A. Panaitescu \\ \hfill }

\affiliation{Space \& Remote Sensing, MS D466, Los Alamos National Laboratory, Los Alamos, NM 87545, USA \\ \hfill}

\begin{abstract}

 We derive basic analytical results for the timing and decay of the GRB-counterpart and delayed-afterglow
light-curves for a brief emission episode from a relativistic surface endowed with angular structure,
consisting of a uniform Core of size $\theta_c$ (Lorentz factor $\Gamma_c$ and surface emissivity $\inu$
are angle-independent) and an axially-symmetric power-law Envelope ($\Gamma \sim \theta^{-g}$).

 In this "\LAE" model, radiation produced during the prompt emission phase (GRB) at angles $\theta > \theta_c$
arrives at observer well after the burst (delayed emission). The dynamical time-range of the very fast-decaying
GRB "tail" and of the flat afterglow "plateau", and the morphology of GRB counterpart/afterglow, are all determined
by two parameters: the Core's parameter $\Gamma_c \theta_c$ and the Envelope's Lorentz factor index $g$,
leading to three types of light-curves that display three post-GRB phases (type 1: tail, plateau/slow-decay,
post-plateau/normal-decay), two post-GRB phases (type 2: tail and fast-decay), or just one (type 3: normal decay).

 Increasing $\Gamma_c \theta_c$ from unity (as for type 3 afterglows) to few/several (as for type 1 afterglows)
opens a gap of duration factor $(\Gamma_c \theta_c)^4$ between the GRB prompt phase and the post-plateau,
where the de-beamed GRB tail and afterglow plateau emissions appear, but the normal decay flux "remembers"
the prompt emission, as the back-extrapolation of the \LAE post-plateau flux should approximately match the GRB's.

 We identify two correlations: GRB tails with a substantial spectral softening or with a larger flux decrease
should be followed by longer afterglow plateaus.
 We show how X-ray light-curve features can be used to determine Core and Envelope dynamical and spectral parameters.
 Testing of the \LAE model is done using the Swift/XRT X-ray emission of two afterglows of type 1 (060607A, 061121),
one of type 2 (061110A), and one of type 3 (061007). We find that the X-ray afterglows with plateaus require an
Envelope Lorentz factor $\Gamma \sim \theta^{-2}$ and a comoving-frame emissivity $\inu \sim \theta^2$, thus, for
a typical afterglow spectrum $F_\nu \sim \nu^{-1}$, the lab-frame energy release is uniform over the emitting surface.

 For two out of three afterglows with optical coverage, we find that the \LAE model can explain only the optical
data during the X-ray normal flux-decay phase.
 Furthermore, the\LAE model can accommodate only optical and X-ray light-curves that are well coupled, displaying
{\sl achromatic} breaks and similar optical and X-ray flux decay indices. Because the \LAE model is more apt to account
for the X-ray afterglow than for the optical, the combination of X-ray radiation from the delayed \LAE and optical
emission from the standard forward-shock could be the most natural explanation for decoupled optical/X-ray afterglows
that display chromatic breaks or plateaus, occurring in only one band.

\end{abstract}

\vspace{3mm}
\section{\bf Introduction}

 Swift/XRT has evidenced the existence of slow-decays/plateaus in the early X-ray afterglow emission (Nousek et al 2006, 
Panaitescu et al 2006a, Zhang et al 2006) following a Gamma-Ray Burst (GRB). The delayed afterglow emission could be 
produced either {\sl after} the burst prompt phase or {\sl during} that phase but arriving later at observer. 

 In the former case, the leading model is the "relativistic blast-wave" (Meszaros \& Rees 1997), where power-law 
afterglow light-curves arise from the synchrotron emission produced at the external forward-shock driven by the 
GRB ejecta into the ambient medium and at the reverse-shock that energizes the incoming ejecta when they arrive 
at the forward-shock. 
 The forward-shock emission from a uniform jet has been used to model the broadband emission (radio, optical, X-ray) 
of dozens of GRB afterglows, beginning with GRB 970508 (Panaitescu, Meszaros, Rees 1998), whose light-curve substantial
brightening suggested that energy is added to the blast-wave. 
 In this model, X-ray plateaus are most naturally attributed to an increase of the shock energy over the region of 
ever-increasing angular extent $\Gamma^{-1}$ that is visible to the observer, $\Gamma$ being the Lorentz factor of 
the decelerating blast-wave. 

 X-ray afterglows could also arise from the same "central-engine" that produces the GRB prompt emission, if it operates
for a longer time. Kumar, Narayan, Johnson (2008) have attributed X-ray plateaus to a variable accretion viscosity 
parameter or to the fall-back of the stellar envelope on the black-hole, thus X-ray counterpart and afterglow 
light-curves could be inverted to infer the accretion-rate and mass fall-back histories.

\vspace{2mm}
\subsection{X-ray plateaus from an increasing blast-wave energy}

 For an outflow where the kinetic-energy per solid-angle $dE_k/d\Omega$ is not uniform (i.e. an outflow endowed with 
angular structure -- Rossi, Lazzati \& Rees 2002), the apparent increase of the blast-wave energy over the visible 
area could be due to the "emergence" of region of higher $dE_k/d\Omega$ at larger angles. For a $dE_k/d\Omega$ that
decreases with angle and an off-axis (or off-uniform Core) observer location, Panaitescu \& Kumar (2003) have shown
that, when the direction toward the observer enters the $\Gamma^{-1}$ relativistic beaming cone of the Core's emission,
the afterglow light-curve displays a phase of slow decay lasting about one decade in time. In addition to being
too short-lived to account for some X-ray light-curve plateaus, the structured outflow post-plateau flux decay is steeper 
than the pre-plateau decay, which could be in contradiction with observations (the steepness of the GRB tail flux 
decay can be substantially reduced if time is measured from the last GRB pulse).

 For an angularly uniform outflow, the increase of the blast-wave kinetic energy could also be due to the ejecta being
sufficiently-wide distributed in radius, the energy increase being produced when new ejecta arrive at the blast-wave
as it decelerates, as in the sustained energy-injection model proposed by Rees \& Meszaros (1998). 
 Panaitescu \& Vestrand (2012) have argued for the likely ubiquity of energy-injection in Swift/XRT afterglows, 
based on the substantial increase in the fraction of X-ray afterglows with late (after 10 ks) light-curve breaks 
(displaying pre- and post-break power-law indices) compatible with the expectations for the forward-shock synchrotron 
emission (from a conical or a spreading jet) when a steady energy-injection is allowed. Enabling energy-injection in 
the blast-wave amounts to introducing one free model parameter (the power-law evolution of the injected energy) for 
two observational constraints (the pre- and post-break flux decay indices), thus the model with a steady energy-injection 
is still over-constrained by observations.

\vspace{2mm}
\subsection{Previous models for chromatic X-ray light-curve breaks}

 Many Swift/XRT afterglows display {\sl chromatic} X-ray light-curve {\sl steepenings} that do not appear in the 
optical (e.g. Panaitescu et al 2006b). Such breaks cannot be explained by a change in the blast-wave dynamics because
any light-curve feature originating in the dynamics of the source must be an {\sl achromatic} feature, appearing at 
all photon energies, even when the reverse-forward shock duality is taken into account, as the dynamics of these 
two shocks are coupled. In the blast-wave model, the existence of two sources of emission could allow for chromatic 
light-curve flattenings when the one shock's emission overtakes the other one's in only one band (e.g. the 10 ks
bump in the optical light-curve of GRB 130427A), but such "over-takings" cannot produce light-curve steepenings.

 Instead, the passage of a spectral break could yield a chromatic light-curve steepening in the band that it 
crosses, with the flux power-law decay index increasing by $\Delta \alpha = \Delta \beta |d\ln \nu_b/d\ln t|$, 
where $\Delta \beta$ is the change in the spectral slope across the break frequency $\nu_b$. The passage of a
spectral break should be accompanied by a spectral softening of $\Delta \beta_{max} \geq 5/6$ if the traveling 
break is the lowest-energy break of the synchrotron spectrum and by $\Delta \beta_{min} \geq 1/2$ for the 
highest-energy spectral break. However, a spectral softening of that magnitude accompanying a light-curve break
is rarely allowed by Swift/XRT measurements, thus this origin for chromatic light-curve breaks is generallya
ruled-out by observations.

 It is possible that afterglow plateaus arise from a reprocessing of the forward-shock emission, such as in the 
"bulk-scattering" model (Panaitescu 2008) where the optical forward-shock emission is up-scattered into the 
X-ray by the incoming (leptonic) ejecta that yield the usual energy-injection in that shock. The upscattered 
emission "reflects" the properties of the scattering ejecta (the radial distribution of their optical thickness 
and Lorentz factor), thus decoupled afterglow light-curves result if the up-scattered emission overshines only 
in the X-ray that coming directly from the forward-shock. Furthermore, dense and geometrically thin reflecting 
sheets in the incoming outflow can explain X-ray flares.

\vspace{2mm}
\subsection{A new model for X-ray light-curve plateaus and chromatic breaks}

 As proposed by Oganesyan et al (2019), the afterglow emission could also be produced during the prompt phase 
by the same mechanism that yields the GRB emission, but arrives later at the observer because it was emitted 
from an angle larger than the $\Gamma^{-1}$ visible during the GRB phase, hence the photon travel path to observer 
is longer. 

 The success of the "reprocessing" and the "central-engine" models stems from the one-to-one correspondence between 
the afterglow X-ray flux and the history of the mechanism that produces that emission. Similarly, in the \LAE model 
there is a bijection between the X-ray flux and two properties of the radiating surface: its local Lorentz factor and
emisivity.

 The purpose of this article is to derive analytically the \LAE counterpart and afterglow timing and flux decay 
indices, to correlate model parameters with light-curve features, and to apply this model to real GRB afterglows, 
with the aim of identifying model features/parameters that explain X-ray plateaus, post-plateaus and, possibly, 
the optical afterglow emission.

\vspace{3mm}
\section{\bf The Simple Formalism of the Large-Angle Emission Model}

 We make the assumption that each GRB pulse is the instantaneous emission from a infinitesimally-thin surface, 
so that the comoving frame duration of the emission due to electron cooling or the radial thickness of the emitting
shell are negligible in comparison with the timescale for photon arrival-time arising from the spherical curvature
of that emitting surface. This approximation is obviously correct at times well after the pulse duration and 
yields pulses that have an infinitely sharp rise.
The great advantage of this approximation is that it allows a one-to-one correspondence 
between the location where emission is produced (the angular offset $\theta$ between the local direction toward 
the observer and the surface origin--observer line-of-sight) and the photon arrival-time at observer $t$.
In principle, {\sl this $\theta-t$ bijection allows one to invert any afterglow X-ray light-curve} and obtain
a constrain a combination of the angular distribution of the Lorentz factor $\Gamma(\theta)$ and the surface 
emissivity $\inu (\theta)$. 

 The {\sl most important assumption is that dynamics and emission from this surface is axially symmetric}, all
quantities being a function of the offset angle $\theta$. A less important approximation is that the observer 
is located on the axis of symmetry, but this is not a very restrictive approximation because the observer should
be biased toward detecting mostly bursts which are seen from within their uniform, bright Core, in which case
only the early afterglow light-curve (arising from the Envelope) would be weakly dependent on the exact location 
of the observer within that Core. 
 
 These two last assumptions afford an analytical treatment to obtain afterglow light-curves and come at the price 
of yielding light-curves that may have not enough diversity to account for real afterglow light-curve features, 
such as short-lived fluctuations (flares). Consequently, numerical best-fits to real afterglows with the \LAE model 
under these two approximations are likely to have large statistics estimators like $\chi^2$, and the quality of 
the best-fit should be validated visually.

\vspace{2mm}
\subsection{\bf Kinematics of the Photon Arrival(-Time)}

 The observer frame photon arrival-time $t_{obs}$ for the emission released by the fluid moving at constant speed $v$, 
at angle $\theta$ relative to the direction toward the observer, and produced at lab-frame time $t_{lab}$ is
\begin{equation}
  t_{obs} (\theta) = t_{lab} [1 - v(\theta) \cos \theta]
\label{tobs}
\end{equation}
with $t_{lab}$ measured since the launch of the radiating surface, and with $v$ in units of the speed of light. 
For an observer on the symmetry axis and for a velocity $v(\theta)$ that does not increase with angle, the first 
arriving photons are those emitted from $\theta=0$, their arrival-time being 
\begin{equation}
 t_o \equiv t_{obs} (0) = t_{lab} [1 - v(0)] \simeq \frac{R_o}{2c\Gamma^2(0)}
\end{equation}
where $R_o \equiv R(0)$ and $\Gamma(0) \gg 1$ are the emission-release radius and the Lorentz factor on the symmetry axis.
In following equations, the observer time is measured from the {\sl the onset/peak time} $t_o$ of the burst last 
pulse, which, owing to the very fast decay of the \LAE, is most likely dominating the afterglow emission, but we note 
that this peak epoch $t_o$ is not measured from the GRB onset (as in figures) because that time is set by the arrival
of photons produced by the first GRB pulse.
Thus
\begin{equation}
  t(\theta) \equiv t_{obs} (\theta) - t_{obs} (0) = t_{lab} [v(0) - v(\theta) \cos \theta]
\end{equation}
For a relativistic surface motion ($\Gamma \gg 1$) and for small angles $\theta \ll 1$, one can substitute
$v \simeq 1 - 1/2\Gamma^2$ and $\cos \theta \simeq 1 - \theta^2/2$, leading to
\begin{equation}
  \frac {2t(\theta)}{t_{lab}} \simeq \theta^2 + \frac{1}{\Gamma^2(\theta)} - \frac{1}{\Gamma^2(0)}
\label{t}
\end{equation}

 Equation (\ref{t}) for the "kinematics" of the \LAE relates the photon arrival-time $t$ with the emission offset 
angle $\theta$ for a given angular-distribution of the surface's "dynamics" $\Gamma(\theta)$, and must be inverted 
to $\theta(t)$ in order to derive \LAE light-curves. We make the assumption that the relativistic surface has a 
{\sl uniform Core} of angular size $\theta_c$, where the Lorentz factor is a constant $\Gamma_c$, and a 
{\sl power-law Envelope} where $\Gamma$ decreases as a power-law in angle
\begin{equation}
  \Gamma (\theta) = \Gamma_c  \left( \frac{\theta}{\theta_c} \right)^{-g} \;, 
              \quad g \sim \left\{ \begin{array}{ll} 0 & \theta < \theta_c \\  > 0 & \theta_c < \theta 
                                  \end{array} \right.
\label{Gamma}
\end{equation}
with $\Gamma_c \equiv \Gamma(0)$.
A uniform Core with a power-law Envelope is simple enough to allow analytical calculations and leads to afterlow
plateaus that are more prominent than a Gaussian $\Gamma(\theta)$. 
 For it, equation (\ref{t}) becomes
\begin{equation}
 X^g + \frac{t_p}{\tg} X = \frac{t}{\tg} + 1 \;, \quad X \equiv \left( \frac{\theta}{\theta_c} \right)^2
\label{x}
\end{equation}
where
\begin{equation}
 \tg \equiv \frac{t_{lab}}{2\Gamma_c^2} = \frac{R_o}{2c\Gamma_c^2} \;, \quad t_p \equiv (\Gamma_c \theta_c)^2 \tg > \tg
\label{tgtp}
\end{equation}
whose meaning will become clear soon. If the Core expands sideways at the speed of sound, then 
$\Gamma_c \theta_c > 1/\sqrt{3}$ is expected.

 Equation (\ref{x}) can be solved analytically for certain values of the index $g$, including the $g=2$ that appears
to be the value compatibls with most Swift/XRT light-curve plateaus, but we continue with the asymptotic cases
$X < 1$ and $X \gg 1$.

\vspace{2mm}
\subsubsection{GRB Emission from Core ($\theta < \theta_c$, $X < 1$)}

 In this case, $g=0$ in equation (\ref{x}) and its solution is
\begin{equation}
 \theta (t) = \Gamma_c^{-1} \left( \frac{t}{\tg} \right)^{1/2} = \theta_c \left( \frac{t}{t_p} \right)^{1/2} , 
     \Gamma (t) = \Gamma_c  \; (GRB)
\label{tGgrb}
\end{equation}

 A quantity of interest is $\Gamma \theta$, which indicates if the emission is relativistically beamed toward the 
observer or away from her 
\begin{equation}
 \Gamma \theta = \Gamma_c \theta_c  \left( \frac{t}{t_p} \right)^{1/2} \hh = \left( \frac{t}{\tg} \right)^{1/2} \hh = 
           \left\{ \begin{array}{ll} < 1 & \hh t < \tg  \\ 1 & \hh t = \tg \\
                                   > 1 & \hh \tg < t \\ \Gamma_c \theta_c > 1 & \hh t = t_p
           \end{array} \right. 
\end{equation}
thus the observer is within the cone of relativistic beaming ($\theta < \Gamma_c^{-1}$ - radiation beamed toward
the observer) until $\tg$ and outside it ($\theta > \Gamma_c^{-1}$ - radiation is "de-beamed" relativistically) 
after that. {\sl The epoch $\tg$ marks} the end of the last dominant pulse (and of the burst) and 
{\sl the beginning of the very fast-decaying GRB tail}. 

 Another quantity of interest is {\sl the Doppler factor}, which determines the relativistic beaming and boost of 
the received emission:
\begin{equation}
 D (\theta) \simeq \frac{2 \Gamma}{1 + \Gamma^2 \theta^2} \simeq \left\{ \begin{array}{ll} 
         2\Gamma  & \theta \ll \Gamma^{-1} \\ 2/(\Gamma \theta^2) & \theta \gg \Gamma^{-1} \end{array} \right.
\end{equation}
thus
\begin{equation}
 D(t) = 2\,\Gamma_c \left\{ \begin{array}{lll} 
        \hh 1 & \hh t \ll \tg & \hh (GRB) \\ \hh  1/2 & \hh t = \tg & \hh (GRB\; end) \\
        \hh (t/\tg)^{-1} & \hh \tg \ll t & \hh (GRB Tail) \\ 
        \hh \left[ (\Gamma_c \theta_c)^2 + 1 \right]^{-1} & \hh t = t_p  & \hh (Plateau\; start) 
         \end{array} \right.
\label{D}
\end{equation}
Given that the light-curve depends strongly on the Doppler factor, this equation also shows that $\tg$ marks
the end of the flat pulse/burst emission (when the relativistic boost is constant) and the beginning of the
very fast-decaying GRB tail (when $D$ decreases).

 Equation (\ref{tGgrb}) shows that $t_p$ corresponds to the observer receiving emission from the edge of the Core, 
which produces the prompt pulse emission. Adding that the emission from the Envelope yields the afterglow plateau, 
it follows that {\sl $t_p$ marks the end of the GRB tail and the beginning of the afterglow plateau}.

%\vspace{2mm}
\subsubsection{Afterglow Emission from Envelope ($\theta > \theta_c$, $X > 1$)}

 Equation (\ref{x}) can be solved in the asymptotic case when one of the terms in the left-hand side
is dominant. Taking into account that $t_p > \tg$ for $\Gamma_c \theta_c > 1$ (equation \ref{tgtp}), 
the second term is dominant at $X \simg 1$ and the approximate solution of equation (\ref{x}) is the same as 
for the Core, thus
\begin{equation}
 \theta (t) = \theta_c \left( \frac{t}{t_p} \right)^{1/2} \;,\;
 \Gamma (t) = \Gamma_c \left( \frac{t}{t_p} \right)^{-g/2} \; (Plateau)
\label{tGp}
\end{equation}
Therefore
\begin{equation}
 \Gamma \theta = \Gamma_c \theta_c \left( \frac{t}{t_p} \right)^{-(g-1)/2} > 1 \quad (Plateau)
\label{Gt}
\end{equation}
is increasing if $g<1$ and is decreasing if $g>1$, remaining above unity (i.e. radiation is de-beamed relativistically) 
until epoch
\begin{equation}
 t_{pp} = \left( \Gamma_c \theta_c \right)^{2/(g-1)} t_p \quad (g>1)
\label{tpp}
\end{equation}
and the Doppler factor for $\theta > \Gamma^{-1}$ is
\begin{equation}
 D(t) \simeq \frac{2}{\Gamma \theta^2} = \frac{2\Gamma_c}{(\Gamma_c \theta_c)^2} \left( \frac{t}{t_p} \right)^{(g-2)/2} 
     \quad (Plateau)
\label{Dp}
\end{equation}

 For $g<1$, the second term in the left-hand side of equation (\ref{x}) is dominant indefinitely and equation 
(\ref{Gt}) shows that $\Gamma \theta$ increases and remains always above unity, thus the plateau phase never ends. 
However, the Doppler factor decreases and the afterglow flux will decrease even faster, thus "plateau" is now 
just a name for the $\Gamma \theta > 1$ phase. 

 Because the afterglow light-curve depends mostly on the Doppler factor evolution, it follows from equation
(\ref{Dp}) that $g < 2$ yields decaying plateaus, $g \simeq 2$ produces flat plateaus, while $g > 2$ leads 
to rising plateaus. Most Swift/XRT plateaus are slowly decaying or flat, which indicates that {\sl \LAE plateaus 
must have $g \simeq 2$}. Then equation (\ref{tpp}) could be used to determine the Core parameter $\Gamma_c \theta_c = 
\sqrt{t_{pp}/t_p}$, but it is hard to determine from light-curve plateaus the plateau end epoch $t_{pp}$ defined by 
$\Gamma \theta = 1$.

 It can be shown that, for $t > t_{pp}$, the first term in the left-hand side of equation (\ref{x}) is dominant,
thus its approximate solution for $g>1$ is
\begin{equation}
 \theta (t) = \theta_c \left( \frac{t}{\tg} \right)^{1/2g}  ,\;
 \Gamma (t) = \Gamma_c \left( \frac{t}{\tg} \right)^{-1/2} (post-Plateau)
\label{tGpp}
\end{equation}
Consequently
\begin{equation}
 \Gamma \theta = \Gamma_c \theta_c \left( \frac{t}{\tg} \right)^{-(g-1)/2g} < 1 \quad (post-Plateau)
\end{equation}
and the Doppler factor for $\theta < \Gamma^{-1}$ is
\begin{equation}
 D(t) \simeq 2\Gamma = 2 \Gamma_c \left( \frac{t}{\tg} \right)^{-1/2} \quad (post-Plateau)
\end{equation}
A decreasing Doppler factor yields a decreasing \LAE afterglow, hence the "post-plateau" phase is the usual 
afterglow decay.

 The equations above show the effect of a Lorentz factor decreasing with angle in the Envelope and the distinction 
between plateau and post-plateau phases. {\sl During the plateau}, the \LAE is in the $\theta > \Gamma^{-1}$ case, 
the emission is relativistically de-beamed, which explains why the plateau flux is much dimmer than the prompt flux
but $\Gamma \theta$ and $\Gamma$ decrease, so the Doppler factor $D \sim \Gamma/(\Gamma\theta)^2$ is about constant, 
yielding a nearly constant flux.
{\sl After the plateau}, the \LAE enters the $\theta < \Gamma^{-1}$ case, the emission is relativistically beamed 
toward the observer but $\Gamma$ and $D \sim \Gamma$ decrease, leading to a fast flux decay (see below).
Consequently, {\sl the epoch $t_{pp}$ given in equation (\ref{tpp}) marks the end of the plateau} and the beginning 
of the normal afterglow decay.

 Upper panels of {\bf Figure 1} show the evolution of $\Gamma \theta$ and of the Doppler factor.
As expected, afterglow plateaus are obtained for $g \simeq 2$ and the epoch ratios satisfy $t_p/\tg \simeq 
t_{pp}/t_p \simeq (\Gamma_c \theta_c)^2$, i.e. GRB tails and plateaus last longer for Cores increasingly wider relative 
to the region producing the GRB prompt emission (of angular size $\Gamma_c^{-1}$).

\vspace{2mm}
\subsection{\bf \LAE Light-Curves}

 The relativistic motion of the emitting surface beames the comoving-frame emissivity $i'(\nu')$ by a factor $D^2$ in 
the direction toward the observer. The flux energy density $F_\nu = dE/d\nu$ remains unchanged after relativistic 
transformation because the increase of photon energy by a factor $D$ ($\nu = D\nu'$)appears both in $dE$ and $d\nu$. 
The contraction of time-intervals by a factor $D$ from comoving to observer frames does not apply if the emission 
is instantaneous, the observer-frame rate being instead determined by the spread the in photon arrival-time due to 
the spherical curvature of the emitting surface, so the received flux should be proportional to $dA/dt$ with 
$dA = 2 \pi R^2 \sin \theta d\theta$ being the infinitesimal area at angles $(\theta,\theta+d\theta)$ corresponding 
to arrival times in $(t,t+dt)$, and corrected by a multiplicative factor $\cos \theta$ because $dA$ is seen at angle
$\theta$ and subtends a solid angle proprtional to $\cos \theta$. 
 Putting these factors together, the received \LAE flux is
\begin{equation}
 F_\nu \sim D^2 i'\left( \frac{\nu}{D} \right) \sin(2\theta) \frac{d\theta}{dt}
\end{equation}

 The spectrum of the comoving-frame surface emissivity is a power-law in photon energy:
\begin{equation}
 \i'(\nu') = i'_p (\theta) \left[ \frac{\nu'}{\nu'_p(\theta)} \right]^\beta 
\end{equation}
with the spectral slope $\beta$ having a fixed value in the Core and in the Envelope (but Swift X-ray measurements
reuire that these two values differ), and with $i'_p$ the flux at the peak-energy $\nu'_p$, 
Assuming that the peak spectral characteristics $i'_p$ and $\nu'_p$ are also simple power-laws in the location angle
$\theta$
\begin{equation}
 i'_p \sim \theta^y \;, \; \nu'_p \sim \theta^z \rightarrow \inu \sim \theta^x \;,\; x \equiv y - \beta z
\label{yz}
\end{equation}
with $y=z=0$ for the uniform Core, the \LAE becomes
\begin{equation}
 F_\nu (t) \sim D^{2-\beta} \frac{d\theta^2}{dt} \theta^x \nu^\beta 
\label{Fnu}
\end{equation}
in the $\theta \ll 1$ limit.
If all observations are in the same power-law branch $\nu^\beta$, then only the index $x$ can be determned. 
However, optical and X-ray observations on both sides of the peak-energy $\nu'_p$, probing both power-law 
spectral branches, break that degeneracy and enable the determination of both indices $y$ and $z$.

 Substituting the time-dependencies of $D$ and $\theta$ derived in previous section, one obtains the \LAE
light-curve decay index
\begin{equation}
 F_\nu (t) \sim t^{-\alpha} \nu^\beta
\label{alfa}
\end{equation}
\begin{displaymath}
 \alpha = \left\{ \begin{array}{lll} 
           0 & \hh t < \tg & \hh  (GRB) \\  2 - \beta & \hh \tg < t < t_p & \hh (Tail) \\
           \frac{1}{2} [(2-g)(2-\beta)-x] & \hh t_p < t < t_{pp} & \hh (Plateau) \\
           \frac{1}{2} [4 - \beta - (x+2)/g ] & \hh t_{pp} < t \;(g>1)  & \hh (post-P) \\
         \end{array} \right.
\end{displaymath}
These flux decay indices are asymptotic values and are displayed by the numerically-calculated light-curves
(lower right panel of Figure 1) only late in the GRB tail ($\tg \ll t \siml t_p$), early in the afterglow plateau 
($t_p \siml t \ll t_{pp}$), and late in the post-plateau decay ($t_{pp} \ll t$). Furthermore, Figure 1 shows that
the first two asymptotic values are reached only if the GRB tail and afterglow plateau are sufficiently long-lived
(or well-developed), as for $\Gamma_c \theta_c \simg 5$.

\vspace{2mm}
\subsubsection{GRB Tail} 
\label{grbtail}

 The very fast decay of the GRB tail given in equation (\ref{alfa}) was first noted by Fenimore, Madras \& Nayakshin 
(1996), re-calculated by Kumar \& Panaitescu (2000), and used by Liang et al (2006) to fit Swift GRB tails.

 That fast decay could be used to constrain the Core parameter $\Gamma_c\theta_c$ in the following way. 
Acording to equation (\ref{D}), from the last pulse peak epock $t_o$ to the beginning of the GRB tail $\tg$,
defined by $\theta = \Gamma^{-1}$, the Doppler factor decreases by a factor 2, thus the pulse flux at the GRB
end epoch $\tg$ is a factor $\siml D^{2-\beta} = 2^{2-\beta}$ lower than at the peak, which allows a rough
determination of $\tg$ from the pulse light-curve. Then equation (\ref{tgtp}) allows one to determine 
$\Gamma_c \theta_c$. However, the assumption of a quasi-instantaneous release of radiation in the \LAE model 
is likely to be inaccurate during the pulse, thus this constraint on $\Gamma_c \theta_c$ may be not so tight.

\vspace{2mm}
\subsubsection{Afterglow Plateau} 
\label{plateau}

 The previously discussed constancy of the Doppler factor $D$ for a Lorentz factor distribution 
$\Gamma \sim \theta^{-2}$ in the Envelope is manifested by the vanishing of the first term in equation (\ref{alfa}), 
and the plateau has an index $\alpha_p=-x/2$ dependent only on the angular distribution of the emission peak spectral 
characteristics in the Envelope: if they are uniform with angle ($x=0$), then the plateau is flat ($\alpha_p=0$). 

 Equation (\ref{alfa}) shows that the emissivity power-law index that leads to an afterglow plateaus from a surface 
of uniform Lorentz factor ($g=0$), is $x = 2(2-\beta-\alpha_p) = 2(2-\beta) \simeq 6$ for a flat plateau ($\alpha_p=0$) 
and a typical spectral slope $\beta \simeq -1$. This illustrates the effectiveness of a decreasing Lorentz factor 
in producing afterglow plateaus.

 The upper right panel of Figure 1 indicates that X-ray plateaus should require an Envelope Lorentz factor index 
$g \simeq 2$. How robust is that conclusion might be assesed from equation (\ref{alfa}) for the plateau decay index: 
a variation $\delta x$ in the unknown index of the emissivity $\inu$ leads to an uncertainty $\delta g$ in the
determination of $g$ from the plateau flux-decay index $\alpha_p$ of $\delta g = \delta x/(2-\beta) < \delta x/3$.
 
 The Envelope index $g$ may be determined better using equation (\ref{tpp}) if the plateau end epoch $t_{pp}$ is 
measured from the X-ray light-curve and if the Core parameter $\Gamma_c\theta_c$ is constrained from the GRB tail 
timing, as described in \S\ref{grbtail}.

\vspace{2mm}
\subsubsection {Post-Plateau Afterglow}

 For the Lorentz factor angular distribution index that yields plateaus ($g=2$), equation (\ref{alfa}) shows that
the post-plateau flux decay index is $\alpha_{pp} = (3-\beta-x/2)/2$ thus, for a typical spectral slope $\beta \simeq -1$, 
the \LAE post-plateau flux should exhibit a decay with index $\alpha_{pp} = 2 -x/4$. Most post-plateaus measured by 
Swift/XRT have a decay index $\alpha_{pp} \simeq 1.5$, which requires an emissivity index $x = 4(2-\alpha_{pp}) \simeq 2$, 
thus the surface emissivity must increase with angle, to mitigate the steep decay of the \LAE caused by the decrease 
of the Doppler factor during the post-plateau phase. 

 Equation (\ref{alfa}) shows that afterglow measurements provide two constraints (the plateau and post-plateau
decay indices) for two Envelope unknowns (Lorentz factor index $g$ and emissivity index $x$) and, if the index $g$
is constrained from the plateau timing (as described in \S\ref{plateau}), then one could test the \LAE model analytically. 
Two caveats apply. First is the difficulty in measuring the GRB end epoch $\tg$ which leads to an inaccurate index $g$.
Second is that the plateau and post-plateau flux decay indices given in equation (\ref{alfa}) are asymptotic values, 
hence their accurate measurement is not trivial. Instead, a more reliable determination of all the \LAE model parameters 
and its testing can be achieved through numerical fits of observations with the \LAE model light-curves.

\vspace{3mm}
\section{\bf Observational features of the Large-Angle Emission}

\vspace{2mm}
\subsection{\bf Spectral Softening during the GRB Tail}

 CGRO/BATSE, Fermi/GBM, Konus-Wind spectra of GRBs show that, during the prompt phase, the observer-frame peak-energy 
$\nu_p = D\nu'_p$ of the Core emission is somewhere in the mid-to-hard X-rays, at 20--500 keV. If $\nu'_p$ is 
angle-independent in the uniform Core, then the decrease of the Doppler factor from $D(t_o)=2\Gamma_c$ for 
$\theta = 0$ to $D(\tg)=\Gamma_c$ for $\theta=\Gamma_c^{-1}$ (equation \ref{D}) implies a decrease by a factor 2 
for the observer peak-energy $\nu_p$ thus, at the end of the pulse, $\nu_p$ should still be in the mid X-rays. 
The spectrum of received emission continues to soften after $\tg$ and the peak-energy $\nu_p$ should decrease by 
another factor $\simeq (\Gamma_c \theta_c)^2/2$ until $t_p$ (equation \ref{D}), when the Core-Envelope boundary becomes 
visible and the afterglow plateau begins. 

 Swift/XRT has evidenced in many cases a spectral softening of the 1-10 keV emission during the last pulse,
with the spectrum having in some cases a positive slope $\beta_o^{(C)}$ at the peak epoch, thus $\nu_p (t_o) > 10$ keV, 
and a slope $\beta_x^{(C)} < -1$ at the end of the GRB tail, thus $\nu_p (\tg) < 1$ keV. Therefore, a significant 
spectral softening with $|\Delta \beta| > 1$ during the GRB tail requires $\Gamma_c \theta_c \simg 3$ and 
a hard low-energy spectral slope $\beta_o^{(C)} = 1/3$ (as for synchrotron emission below the spectrum peak),
with a lower value for $\Gamma_c\theta_c \in (1,2)$ being possible for bursts with a lesser spectral softening 
$|\Delta \beta| < 1$ during the GRB tail. In the latter case, the smaller is the Core factor $\Gamma_c\theta_c$, 
the more likely that a hard low-energy slope is needed to acquire the spectral softening measured during the GRB tail.

 We note that the above conclusion is quite robust because the decrease of the Doppler factor (and observer-frame 
spectral peak-energy) by a factor $\Gamma_c^2 \theta_c^2$ from pulse peak to end of GRB tail represents an upper 
limit to the possible softening for other angular distributions of the Core Lorentz factor, if it decreases with angle.
As an example, if the Core had a Gaussian Lorentz factor profile -- $\Gamma (\theta) = \Gamma_c \exp(-\theta^2/n\theta_c^2)$ -- 
then the Doppler factor at $\theta_c$ would be a factor $(\Gamma_c\theta_c)^2/e^{1/n}$ smaller than on axis.

\vspace{2mm}
\subsection{\bf Correlations for the \LAE Model}
\label{correl}

 Equations (\ref{tgtp}) and (\ref{tpp}) imply that 
\begin{equation}
 \frac{t_p-t_o}{\tg-t_o} = \left( \frac{t_{pp}}{t_p-t_o} \right)^{g-1} = (\Gamma_c \theta_c)^2 
\label{ttt}
\end{equation}
for afterglows with plateaus, with the pulse beginning/peak epoch $t_o$ being comparable to the GRB end epoch $\tg$, 
likely negligible compared to the plateau beginning epoch $t_p$, and very likely much less than the plateau end time 
$t_{pp}$. 

 Previous considerations about the X-ray spectral softening, that afterglow plateaus likely require $g \geq 2$,
and the second equality given in equation (\ref{ttt}), imply that GRB tails that display a significant softening 
(e.g. from a spectral slope $\beta \simeq 1/3$ to $\beta \siml -1$) should be followed by afterglow plateaus that
last at least a decade time. In general, the GRB tail softening should correlate with the plateau duration because
of their dependence on the Core parameter $\Gamma_c \theta_c$ but this correlation may be weakened by that Swift/XRT 
1-10 keV observations set only a lower limit on the decrease of the peak-energy, thus the full softening of some 
GRB tails may be under-measured. 

 Equation (\ref{alfa}) indicates, that during the GRB tail, from $\tg$ to $t_p$, the X-ray flux should decrease
by a factor $(t_p/\tg)^{2-\beta}$. Adding the first equality of equation (\ref{ttt}), this means that the 
burst-to-plateau flux-ratio should be
\begin{equation}
 \frac{F_{grb}}{F_{pl}} > \left( \frac{t_p}{\tg} \right)^{2-\beta} =
  \left( \frac{t_{pp}}{t_p} \right)^{(2-\beta)(g-1)} 
\end{equation}
Therefore, if the X-ray afterglow plateau is the \LAE, then there should be a correlation between the flux decrease
during the GRB tail and the plateau duration, with larger GRB flux drops being followed by longer plateaus.

\vspace{2mm}
\subsection{\bf Optical Afterglow Brightness}

 From equation (\ref{alfa}), the X-ray and optical flux decay indices during the post-plateau afterglow satisfy
\begin{equation}
 \alpha_x - \alpha_o = \frac{1}{2} (\beta_o-\beta_x) \left\{ \begin{array}{ll}
      \hh 2-g+z, & \hh g < 1 \\ \hh 1-z/g, & \hh g > 1 \end{array} \right. (post-P)
\label{daox}
\end{equation}
For $g < 1$, the post-plateau does not exist formally and the afterglow flux decay is given by the third branch of 
equation (\ref{alfa}). Thus, for a uniform peak-energy $\nu_p^{(E)}$ (i.e. $z=0$) located between optical and X-rays,
one expects an optical flux decay index "slower" than in the X-ray: $\alpha_x - \alpha_o = (\beta_o -\beta_x)/2$ 
for $g>1$ and a value larger by a factor $2-g$ for $g<1$.

 Below the Envelope peak-energy $\nu_p^{(E)}$, the spectral slope could have two values. One is the slope 
$\beta_o^{(E)} = 1/3$ expected for optically thin emission from {\sl uncooled} electrons. In this case, for a 
typical X-ray plateau flux of 0.1-10 mJy and an Envelope peak-energy $\nu_p^{(E)} (t_p) \siml 1$ keV, the \LAE 
afterglow would have an optical magnitude $R=14-19$. For a typical X-ray spectral slope $\beta_x^{(E)} \simeq -1$, 
the optical post-plateau light-curve would decay significantly slower than at X-rays: $\alpha_x - \alpha_o = 2/3$.

 However, it is unlikely that the \LAE from uncooled electrons produces the optical afterglow because the implied
spectral slope $\beta_o^{(E)}=1/3$ is significantly harder than that usually measured, $-1.0 < \beta_o < -0.5)$,
and such a spectral softening cannot be always attributed to just the right amount of dust reddening in the host galaxy.
The relative dimness of the optical afterglow in this case suggests that this hard optical \LAE is rarely observed
because it is likely overshined by the forward-shock emission.

 If the Envelope emission were synchrotron from {\sl cooled} electrons, then its low-energy spectral slope 
would be $\beta_o^{(E)}=-1/2$, more consistent with that typically measured for optical afterglows.
In this case, the optical flux would be a factor $(1\,keV/2\,eV)^{1/2}=20$ brighter than in the X-ray, 
reaching magnitude $R=10-15$. Thus, the \LAE is more likely to explain the brighter optical afterglows, 
provided that its low-energy spectrum has a slope of $\beta_o = -1/2$. 
In this case, the optical and X-ray flux decays are more similar, with $\alpha_x - \alpha_o = 1/4$.

\vspace{2mm}
\subsection{\bf Counterpart-Afterglow Diversity} 

 The various types of afterglows arising from the \LAE from an emitting surface with a uniform Core and a power-law
Envelope can be identified from equation (\ref{Gt}), which shows that the factor $\Gamma\theta$ for the relativistic
beaming rises to $\Gamma_c \theta_c$ (as shown in left upper panel of Figure 1) at the Core-Envelope boundary, 
after which it increases if $g < 1$ or decreases if $g > 1$.

{\sl Type 1} : three post-GRB phases - counterparts with tails and afterglows with plateaus and post-plateaus for
 $\Gamma_c \theta_c > 1$ and $g>1$. \\
For $\Gamma_c \theta_c > 1$, the existence of the GRB tail phase ($\Gamma_c^{-1} < \theta < \theta_c$) and of the 
afterglow plateau ($\Gamma \theta > 1$ at $\theta > \theta_c$) are ensured. For $g > 1$, $\Gamma\theta$ decreases
during the plateau and the post-plateau starts when $\Gamma \theta = 1$ (equation \ref{tpp}). 
In this case, the GRB counterpart and afterglow display all phases: GRB tail, afterglow plateau, afterglow 
post-plateau. 

{\sl Type 2} : two post-GRB phases - counterparts with tails and afterglows without plateaus for $\Gamma_c \theta_c > 1$ 
and $g<1$. \\
 For $g < 1$, $\Gamma \theta$ continues to increase during the plateau phase and, if $\Gamma_c \theta_c > 1$,
then $\Gamma \theta > 1$, i.e. the plateau phase does not end. However, when $g < 1$, the afterglow flux decay 
during the plateau is not slow, thus the GRB tail is followed by a normal afterglow decay. 

{\sl Type 3} : one post-GRB phase - counterparts without tails and afterglows without plateaus for 
$\Gamma_c \theta_c \simeq 1$. \\
 If $\Gamma_c \theta_c \simeq 1$, then equation (\ref{tgtp}) implies that $\tg = t_p$, i.e. the GRB tail 
phase (defined by $\Gamma_c^{-1} < \theta < \theta_c$) does not exist. If $g > 1$, then equation (\ref{Gt}) 
shows that $\Gamma\theta < 1$ for the Envelope emission, thus the plateau phase (defined by $\Gamma\theta > 1$) 
does not exist. The same is shown by equation (\ref{tpp}) which, for $g>1$, implies that $t_p = t_{pp}$.
If $g < 1$, then $\Gamma\theta > 1$ for the Envelope emission at all times, thus the post-plateau phase 
(defined by $\Gamma\theta < 1$) does not exist but, for $g<1$, the plateau flux decay is fast, so that the GRB
phase is followed by a normal afterglow light-curve decay.

 Lower panels of Figure 1 show light-curves of these three types, for comoving-frame peak spectral parameters
(energy and flux) that are uniform in the Envelope. The light-curves of type 3 afterglows is weakly dependent on
the actual value of the Lorentz factor index $g$.

\vspace{2mm}
\subsection{\bf Funny feature of type 1 afterglows} 
\label{funny}

 As indicated by equation (\ref{ttt}) and as discussed above, the duration of the GRB tail and afterglow plateau 
are correlated with with the Core parameter $\Gamma_c \theta_c$. For $\Gamma_c\theta_c = 1$ (type 3 afterglows)
all emission that is relativistically de-beamed relative to the axial observer (the tail and plateau) is missing 
and the emission that is beamed toward the observer is continuous at the burst-afterglow transition, i.e. the Envelope 
flux at the beginning of the normal afterglow decay (which can be seen as a post-plateau decay) is close to the 
Core flux at the GRB end, {\sl if} the peak spectral quantities are continuous at the Core-Envelope boundary. 
A small flux jump is introduced by the discontinuous Core and Envelope high-energy spectral slopes $\beta_C$ and 
$\beta_E$ implied by the spectral hardening measured by Swift/XRT at the plateau beginning. 
 
 As $\Gamma_c \theta_c$ is increased above unity (i.e. going from type 3 to type 1 afterglows), a gap of a duration
factor $(\Gamma_c \theta_c)^4$ appears between the GRB prompt phase and the post-plateau start, filled with the dimmer, 
de-beamed emission of the GRB tail and of the afterglow plateau, but the normal afterglow decay retains a memory of 
the GRB flux: the extrapolation of the post-plateau flux to the burst end should match within 1 dex the flux at 
the GRB end. 

 This can be proven using the flux decay indices of equation (\ref{alfa}) and equation (\ref{ttt}) for the afterglow
timing
\begin{equation}
 \frac{F_x^{(ex)}(\tg)}{F_x(\tg)} = \left( \frac{\tg}{t_p} \right)^{\h \alpha_t}
    \hh \left( \frac{t_p}{t_{pp}} \right)^{\h \alpha_p} 
    \hh \left( \frac{\tg}{t_{pp}} \right)^{\h -\alpha_{pp}} \hh =
    (\Gamma_c \theta_c)^{2(1+\beta_C-\beta_E)}
\label{extr}
\end{equation}
Miraculously, the result above is independent of all Envelope quantities that determine the flux decay indices of 
equation (\ref{alfa}): the Lorentz factor distribution index $g$, the emissivity distribution index $x$, 
and would also be independent of the Core and Envelope spectral slopes $\beta$ if they were the same.

 It is important to note that the GRB end epoch $\tg$ (to where the flux is extrapolated) and the plateau beginning $t_{pp}$ 
(from where the back-extrapolation starts) may not be easily measurable and that, due to the smooth transition from
plateau to the post-plateau, the extrapolation given in equation (\ref{extr}) falls below the extrapolation of the 
asymptotic post-plateau light-curve. That shortcoming may be compensated (under or over) by that the average GRB flux 
is a factor up to $2^{2-\beta_C}$ larger than the flux at $\tg$, thus the extrapolation of the post-plateau asymptotic  
light-curve to the end of the last GRB pulse should be larger than the measured flux by a factor $\sim \Gamma_c^2\theta_c^2$
(Figure 1, lower-left panel). Thus, equation (\ref{extr}) provides an estimation of the Core parameter $\Gamma_c \theta_c$.

\vspace{3mm}
\section{\bf Application to GRB Counterparts and Afterglows}

 Figures 2-5 illustrate the ability of the \LAE model to explain the multiwavelength measurements of the prompt 
(counterpart) and delayed (afterglow) light-curves for the above three types of GRBs.

\vspace{2mm}
\subsection{\bf Type 1: GRB 061121 - Burst with Tail and Afterglow with Chromatic Plateaus}

 The tail of this burst ({\bf Figure 2}) displays a strong softening, from $\beta_o \simeq 1/3$ (as for synchrotron 
emission from {\sl uncooled electrons} below the peak-energy) at the last pulse peak to $\beta_x=-1.4$ before the 
plateau begins, indicating that the peak-energy of the Core evolves from above 10 keV at the pulse peak to less than 
1 keV at the end of the GRB tail. 

 The simultaneity of the optical and X-ray peaks suggests a common origin of both emissions; however, the hard
low-energy slope of the Core emission, with $\beta_o = 1/3$, required by Swift/XRT measurements at the peak
of the last GRB pulse, implies that the optical peak emission should be a factor $(1\,keV/2\,eV)^{1/3}=6$ dimmer
than in the X-ray, while observations show only a factor 2. The optical peak being a magnitude brighter than 
expected for the \LAE model, suggests that there is another mechanism that overshines the \LAE in the optical.

 The \LAE can account for the post-plateau optical afterglow, starting from 500 s, if the Envelope emission has 
the same low-energy slope $\beta_o^{(E)}=1/3$ as the Core, but then the optical afterglow would be much harder 
than measured by Page et al (2007) at 1 ks, thus a significant dust-redenning by the host galaxy is required. 
An Envelope with the softer low-energy spectral slope $\beta_o^{(E)}=-1/2$ expected for synchrotron emission 
from cooled electrons would be more compatible with observations and can account for the optical plateau being 
brighter than in the X-rays, but a good fit to the entire optical light-curve cannot be obtained because the 
their flux decay indices differ by more than expected (1/4). 

 The strongest indication that the \LAE model cannot explain all optical and X-ray data is provided by the 
decoupling of its light-curves, with the optical and X-ray afterglows displaying chromatic 
plateaus, starting at different times. In contrast, the \LAE model can only yield achromatic 
plateaus, occurring at the same time at all energies because these plateaus originate in the Lorentz factora
distribution on the Envelope surface.

 In conclusion, the \LAE model can account well for the X-ray counterpart and afterglow of GRB 061121, 
under-produces optical emission during the counterpart and afterglow plateau, can explain the post-plateau 
optical afterglow, but the chromatic plateaus argue that the entire optical afterglow arises from another 
mechanism, such as the forward-shock (as proposed by Oganesyan et al 2019).

\vspace{2mm}
\subsection{\bf Type 1: GRB 060607A - Burst with Chromatic Tail and Afterglow with Fast Post-Plateau Decay}

 As shown in {\bf Figure 3}, this counterpart displays a tail (of 0.7 decades in time) significantly shorter than 
its long-lived afterglow plateau (of 2 dex, at first sight), that does not satisfy the second correlation in \S\ref{correl}.
 The exact GRB tail duration factor, $(t_p-t_o)/(\tg-t_o)$, depends on the GRB end epoch $\tg$ that cannot estimated
accurately. Nevertheless, the pulse duration $\tg-t_o$ defined by a flux decrease by a factor $\simeq 2^{2-\beta}$ 
is clearly larger than 10 seconds, hence the GRB tail duration factor is less than 10. The plateau and GRB tail 
duration factors can be reconciled if the index $g$ is lower than the "plateau value" $g=2$. Numerically, we obtain 
$g \simeq 1.3$. 

 To compensate for the faster plateau flux decay resulting for this index, the Envelope emissivity must increase 
with angle as $\inu \sim \theta^x$ with $x = 1.5 \div 3$.
 However, such an emissivity distribution leads to a post-plateau X-ray flux decay that is much slower than observed. 
As can be shown using the asymptotic flux decay indices given in equation (\ref{alfa}), there is no combination of 
the Lorentz factor index $g>0$ and emissivity index $x$ that can explain both the plateau and post-plateau flux decays 
measured for 060607A. 

 Consequently, the fast post-plateau X-ray flux decay requires a change in the angular dependence of the surface 
emissivity at the start of the fast X-ray flux decay (~10 ks) or the emitting surface has an opening half-angle
equal to the offset angle $\theta(10\,ks)$, the lack of emitting fluid at larger angles allowing us to see how
how fast turns off the emission from angle $\theta(10\,ks)$. 

 The optical light-curve of this afterglow is decoupled from the X-ray, and does not display a break at the beginning
of the X-ray plateau. Furthermore and fortunately, even the brightest optical flux that the \LAE could produce,
for the Envelope peak spectral parameters constrained by the spectral hardening measured during the plateau, falls
short of measurements. Either of these two facts indicate that the optical emission of 060607A does not originate 
in the \LAE, which makes a strong case for an external shock optical emission even stronger than for 061121.

\vspace{2mm}
\subsection{\bf Type 2: GRB 061110A - Burst with Tail and Afterglow without Plateau}

 {\bf Figure 4} shows a GRB with a tail (thus $\Gamma_c \theta_c > 1$) followed by a normal decay afterglow, 
which suggests an Envelope index $g<1$. In this case, $\Gamma \theta > 1$ during the afterglow phase, always increasing
(equation \ref{Gt} and Figure 1), radiation being beamed away from the observer, as for the plateau phase of afterglows
of type 1. However, owing to the fast decrease of the Doppler factor for $g < 1$, the afterglow light-curve displays a 
normal decay and not a plateau.

 Such afterglows with a normal decay after the bursts are in minority in the Swift/XRT catalog, which suggests that 
Envelopes with index $g<1$ occurs rarely or that there is an observational bias against them. The defining characteristic 
of type 2 afterglows is not necessarily their fast post-burst decay because that decay may be mitigated by an Envelope 
surface emissivity that increases with angle. Instead, the never-ending $\Gamma\theta > 1$ phase for $g<1$ that
implies the lack of an afterglow light-curve break is what should define type 2 afterglows. Consequently, their
identification requires a Swift/XRT long-monitoring. However, owing to their fast decay, afterglows of type 2 may
become undetectable too quickly (over 1 decade in time) to evidence their single power-law decay.

\vspace{2mm}
\subsection{\bf Type 3: GRB 061007 - Burst without Tail and Afterglow without Plateau}

 Owing to the disappearance of the GRB tail, a Core with $\Gamma_c\theta_c=1$ must lead to a burst with a modest
spectral softening because, during the prompt phase ($t_o=\tg$), the Doppler decreases by only a factor 2 
(equation \ref{D}), thus the peak energy of the Core spectrum can cross only a fraction of the 1-10 keV band.
As shown in {\bf Figure 5}, this burst displays a slightly stronger softening than what can be accommodated by 
the \LAE model for $\Gamma_c\theta_c=1$, which suggests that the Lorentz factor is not uniform in the Core, 
but decreases.

 The optical and X-ray afterglow light-curves of GRB 061007 are well coupled, in the sense that optical-to-X-ray 
flux-ratio appears to be nearly constant. Because the optical emission is a factor 100 brighter than in the X-ray, 
an Envelope emission spectrum with a low-energy spectral slope $\beta_o^{(E)} =-1/2$ (corresponding to synchrotron 
emission from cooled electrons) is favored.
 
 The afterglow optical and X-ray flux decay indices cannot constrain well three Envelope indices for the angular 
distribution of the Lorentz factor ($g$) and of the two spectral peak characteristics (peak energy and peak flux -
$y$ and $z$) or, alternatively, the optical and X-ray distribution indices of the surface emissivity ($x$) of
equation (\ref{yz}).

 In contrast with 061007 and 061121, the \LAE model can accommodate both the X-ray and optical afterglow lightcurves 
of GRB 061007.

\vspace{2mm}
\subsection{\bf \LAE Energetics and Constraints on Core Parameters}

 Using equation (\ref{yz}), the energy per unit area of the radiation released by the Envelope is
\begin{equation}
 \frac{dE}{dA} \sim i_p \nu_p \sim \Gamma i'_p \nu'_p \sim \theta^{y+z-g} \stackrel{\beta=-1}{=} \theta^{x-g}
\label{dEdA}
\end{equation}
where a typical X-ray spectral slope $\beta=-1$ allows the use of the emissivity index $x$, which can be determined
from fitting the light-curves of X-ray afterglow with plateaus.

 The energy radiated by the Envelope from angle $\theta(t_p)=\theta_c$ to some location 
$\theta(t>t_p)$ satisfies
\begin{displaymath}
 \frac{ E_{env}(\theta) }{E_{core}} = 
         \frac{ \int_{\theta_c}^{\theta} d\theta\; \theta \left(\frac{dE}{dA}\right)_{env} }
               { \int_0^{\theta_c} d\theta\; \theta \left(\frac{dE}{dA}\right)_{core} } 
\end{displaymath}
\begin{equation}
  \simeq \frac{ \int_{\theta_c}^{\theta} d\theta\; \theta\; [\Gamma i'_p \nu'_p]_{env}(\theta_c) 
             (\theta/\theta_c)^{x-g} } { \int_0^{\theta_c} d\theta \; \theta \; [\Gamma i'_p \nu'_p]_{core} } 
        \simeq \left( \frac{\theta}{\theta_c} \right)^{2+x-g} 
\end{equation}
if the Envelope and Core spectral peak parameters are continuous at the boundary 
($i_p^{(E)}(\theta_c)=i_p^{(C)}$ and $\nu_p^{(E)}(\theta_c)=\nu_p^{(C)}$),
with $E_{core}$ being the energy emitted by the uniform Core during the last dominant GRB pulse. 

 Using equation (\ref{tGgrb}), the energy released by the Envelope until the plateau end (i.e. up to a location angle
$\theta(t_{pp})$) is
\begin{equation}
  \frac{ E_{env}(t_{pp})}{E_{core}} = \left( \frac{t_{pp}}{t_p} \right)^{(2+x-g)/2} \stackrel{x=g}{=} \frac{t_{pp}}{t_p} 
   \quad (Plateau)
\label{E1}
\end{equation}
 For an Envelope with Lorentz factor index $g<1$ (type 2 afterglow), equation (\ref{E1}) gives the Envelope output 
until the last measurement epoch $t_{last}$ (i.e. the energy released by the Envelope up to an angle $\theta(t_{last})$) 
by replacing $t_{pp}$ with that epoch.

 Using equations (\ref{tGp}) and (\ref{tGpp}), it can be shown that the energy released by the Envelope during the 
plateau and post-plateau phases, and received until the last measurement epoch $t_{last}$ satisfies 
\begin{equation}
 \frac{E_{env}(t_{last}>t_{pp})}{E_{core}} = \left( \frac{t_{last}}{\tg} \right)^{(2+x-g)/2g} 
    \stackrel{x=g}{=} \left( \frac{t_{last}}{\tg} \right)^{1/g} 
\label{E2}
\end{equation}
This result also applies for $\Gamma_c \theta_c = 1$ if $g>1$ (type 3 afterglow).

 The last equalities in equations (\ref{E1}) and (\ref{E2}) applies for $g=x$, as could be the case for a typical 
afterglow with a flat plateau ($g=2$) and a post-plateau flux decay index $\alpha_{pp} = 1.5$, and show the linear 
increase of the Envelope radiative output with observation epoch during the plateau, when the Envelope releases 1-2 
orders of magnitude more energy than the Core, and the slower increase of that energy release during the
post-plateau phase, with another 0.5-1 dex increase being accumulated until the last observation epoch.

 The Core output during the last pulse is $E_{core} = (E_{iso}/4\pi) \pi \theta_c^2$, with $E_{iso}$ the 
isotropic-equivalent GRB output during that last pulse, calculated from its measured fluence and the burst redshift.
Thus, equations (\ref{E1}) or (\ref{E2}) allow the calculation of the energy radiated by the Envelope radiation 
until the last X-ray measurement as a function of the Core half-opening angle $\theta_c$. Imposing an upper limit 
$E_{max}$ on the Envelope released energy leads to an upper limit on the Core size $\theta_c$: 
\begin{equation}
 \theta_c < \theta_{c,max} \equiv 2\; \left(  \frac{E_{max}}{E_{iso}} \right)^{1/2} \left\{ \begin{array}{ll} 
          \hh \left( \frac{\ds t_p}{\ds t_{last}} \right)^{(2+x-g)/4}     & (g<1) \\
          \hh \left( \frac{\ds \tg}{\ds t_{last}} \right)^{(2+x-g)/4g}    & (g>1)
      \end{array} \right.
\label{thc}
\end{equation}

  A lower limit on the Core Lorentz factor is set by the Core parameter $\Gamma_c \theta_c$ determined
through modeling the X-ray light-curve and by the upper limit on the Core angular size in equation (\ref{thc}):
\begin{equation}
 \Gamma_c > \Gamma_{min} \equiv \frac{\Gamma_c \theta_c}{\theta_{c,max}}
\end{equation}
 Furthermore, the condition of relativistic motion at the largest offset location, $\Gamma[\theta(t_{last})]
= \Gamma_c [\theta(t_{last})/\theta_c]^{-g} > 1$, also leads to a lower limit on the Core Lorentz factor, 
but this is just a working condition, as the Core does not really care about the Envelope becoming semi-relativistic.
 Lastly, equation (\ref{tgtp}) can be used to set a lower limit on the radius $R_o$ where the emission is produced
\begin{equation}
 R_o > \frac{\ds 2c \tg}{\ds z+1} \Gamma^2_{min} 
\label{R}
\end{equation}

 Equations (\ref{thc})-(\ref{R}) can be used to constrain Core parameters provided that the Lorentz factor and
emissivity distribution indices $g$ and $x$ are well-determined, which can be done only for aferglows with plateaus, 
where the two measurable X-ray flux decay indices provide enough constraints. For afterglows without plateaus, 
the only measurable flux decay index is insufficient to determine both indices $g$ and $x$, not even when optical 
data are included because the X-ray and optical emissivity indices $x$ are different.

 For an upper limit $E_{max}$ on the Envelope energy output of 1\% of the rest-energy of a solar mass, we find the 
following constraints for the two afterglows with plateaus modeled in this work (for which the Lorentz factor index 
$g$ and the emissivity index $x$ can be well determined): \\
1) 061121  ($g=2.0, y+z=1.9, E_{iso}=3.10^{53}$ erg, $\tg=10$ s, $t_{last}=2$ Ms): 
  $\theta_c < 1.7\deg, \Gamma_c > 110$ (but $\Gamma_c > 450$ from working condition $\Gamma (t_{last}) > 1$) 
  and $R_o > 5.10^{16}$ cm),  \\
2) 060607A ($g=1.3, y+z=3.7, E_{iso}=10^{53}$ erg, $\tg=50$ s, $t_{last}=10$ ks): 
  $\theta_c < 0.75\deg, \Gamma_c > 130, R_o > 10^{16}$ cm. 

 For afterglows without plateaus, the indices $g$ and $x$ cannot be constrained with only two flux decay indices 
(X-ray and optical). Assuming $g=2$ for consistency with the index determined for afterglows with plateaus, 
the resulting constraints are: \\
3) 061007 ($g=2, y+z=1.2, E_{iso}=10^{54}$ erg, $\tg=17$ s, $t_{last}=600$ ks): 
  $\theta_c < 4.6\deg, \Gamma_c > 200$ (from $\Gamma (t_{last}) > 1$), $R_o > 2.10^{16}$ cm.

\vspace{3mm}
\section{\bf Conclusions}

\subsection{How does the \LAE model work ?}

 One defining characteristic of the \LAE model considered here is that counterpart and afterglow emissions are 
produced simultaneously over a short time (in the observer frame) but the latter is delayed and received over a 
longer time, owing to the different path length to the observer. We have considered that surface that produces
the burst and afterglow emission has a {\sl uniform Core} of constant Lorentz factor $\Gamma_c$ and co-moving 
frame emissivity $\inu$ and a {\sl power-law Envelope} where these two basic quantities have a power-law 
dependence on the location angle $\theta$. In this \LAE model, the prompt/counterpart emission arises from the Core, 
while the delayed/afterglow radiation is produced by the Envelope.

 In principle, the plateau (slow-decay phase) evidenced by Swift/XRT in the light-curves of most GRB afterglows
could be explained using only the Envelope emissivity's angular distribution, while assuming a radiating surface 
with a uniform Lorentz factor. Then, a surface brightness that increases sharply with angle, as $\inu \sim \theta^6$, 
would be required by flat plateaus.

 However, as noted by Oganesyan et al (2019), an Envelope whose Lorentz factor decreases with angle leads naturally 
to an afterglow plateau when the emission from the Envelope becomes visible to the observer, owing to the slow 
evolution of the Doppler factor $D \sim (\Gamma\theta^2)^{-1}$ when radiation is de-beamed relativistically 
(i.e. when $\Gamma\theta > 1$). This natural accounting of afterglow plateaus by a power-law Envelope is a second
defining characteristic of the \LAE model.

\vspace*{1mm}
\subsection{\sl How does the \LAE explain the counterpart-afterglow diversity ?}

 The morphology of counterparts and afterglows is controlled by two parameters: the uniform Core parameter 
$\Gamma_c \theta_c$ (with $\theta_c$ being its half-opening angle) and the Envelope's index $g$. 
For $\Gamma_c \theta_c > 1$ and $g>1$, the counterpart should display a fast-decaying tail and the afterglow should 
show a slow-decaying plateau (where $\Gamma \theta > 1$) and a post-plateau/normal flux decay (where $\Gamma \theta < 1$), 
as for GRBs 060607A and 061121.
For $\Gamma_c \theta_c > 1$ and $g<1$, the $\Gamma \theta > 1$ phase never ends but, owing to that $g < 1$, the afterglow 
flux decays fast and the light-curve does not display a plateau, as for GRB 061110A. 
If $\Gamma_c \theta_c \simeq 1$, then the GRB tail also disappears and the burst should be followed by a normal afterglow, 
as for GRB 061007.

 From the dependence of the GRB tail and of the afterglow plateau durations on the Core $\Gamma_c\theta_c$ 
(although unexpected at first sight, the Envelope plateau duration cares about a Core parameter - equation \ref{tpp}), 
we have identified two correlations among counterpart and afterglow features: GRB tails that display a stronger 
spectral softening or a larger flux drop should be followed by dynamically longer plateaus.

\vspace*{1mm}
\subsection{\sl How can Core and Envelope parameters be determined ?}

 The spectral characteristics of the uniform Core emission (peak energy, peak flux, spectral slopes below and 
above the peak energy) can be determined by fitting the X-ray counterpart light-curves at 1 keV and 10 keV,
which is the equivalent of fitting the 1-10 keV flux and spectral slope measured by Swift/XRT. 
Evidently, there is some degeneracy among those four parameters, but much of it should/could be removed by modeling
the spectral softening evidenced by XRT during most GRB tails, which proves that the Core peak-energy crosses the
1-10 keV band at that time. 

 X-ray counterpart and afterglow timing observations provide two constraints (the ratios of the GRB end $\tg$, 
the plateau start $t_p$, and the plateau end $t_{pp}$ epochs) for two parameters of the \LAE model: the Core 
parameter $\Gamma_c \theta_c$ and the power-law exponent of the Envelope Lorentz factor distribution $\Gamma \sim 
\theta^{-g}$ (equation \ref{ttt}). The X-ray afterglow plateau and post-plateau flux decay indices $\alpha_p$ and
$\alpha_{pp}$ provide two other constraints for two model parameters: the index $g$ and the exponent of the power-law 
distribution of the surface emissivity $\inu \sim \theta^x$ (equation \ref{alfa}). Thus, X-ray counterpart 
and afterglow observations provide four constraints for three fundamental model parameters ($\Gamma_c\theta_c, g, x$),
which means that the \LAE model with a power-law Envelope is {\sl testable} using afterglows with plateaus, 
albeit that testing can be done only numerically because the GRB end epoch $\tg$, the plateau end epoch $t_{pp}$, 
and the plateau flux decay asymptotic index are difficult to measure directly from light-curves.

 If the peak-energy of the Envelope emission spectrum remains below the X-rays, then fits to only X-ray afterglow 
light-curves cannot separate the degenerate dependence of the surface emissivity on the spectrum peak-energy and 
peak-flux. However, adding optical observations leads to two energies straddling the Envelope peak-energy, which
can break that degeneracy. In fortunate cases, such as that of GRB 060607A, an afterglow spectral hardening indicates
that the Envelope peak-energy increases and crosses the 1-10 keV band, which allows the determination of the Envelope
peak spectral characteristics.

 For the Envelope emission, an analytical argument and numerical fits show that afterglow plateaus require a Lorentz
factor distribution $\Gamma \sim \theta^{-2}$ and that post-plateau X-ray flux decays require an emissivity 
distribution $\inu \sim \theta^2$, which imply that, for a typical X-ray afterglow spectrum $F_\nu \sim \nu^{-1}$, 
the Envelope bolometric surface emissivity is nearly uniform (equation \ref{dEdA}). 

 All milestones of the \LAE light-curve depend on the product $\Gamma_c \theta_c$, and not on those parameters 
individually, thus light-curve fitting cannot break their degeneracy. Instead, an upper limit on the energy of 
the radiation released by the Envelope, together with the measured burst fluence, can be used to set an upper 
limit on the Core half-opening $\theta_c$. For two afterglows with plateaus, 060607A and 061121, we find that
$\theta_c \siml 1\deg$ and that the last GRB pulse emission is produced at $R > 10^{16}$ cm.

\vspace*{1mm}
\subsection{\sl Decoupled afterglow light-curves and the case for the \LAE}

 Inclusion of optical data in the \LAE modeling of X-ray light-curves is warranted only if the optical and X-ray 
light-curves are coupled, i.e. if they track each other fairly well, because the \LAE model (just as the alternate 
external-shock model) cannot produce decoupled afterglow light-curves with chromatic breaks that appear at only 
one observing energy. 
 For example, the optical and X-ray light-curves of GRBs 060607A and 061121 are decoupled, as the GRB tail and 
X-ray plateau of the former and the plateaus of the latter are chromatic. Because of this limitation, the \LAE 
model can explain only a subset of the optical data, mostly during the post-plateau phase.

 Maybe it is worth noting that restricting the use of optical data only to well-coupled light-curves (optical and 
X-ray light-curves with similar decay indisces) implies a 
bias toward selecting only bright optical counterparts and afterglows, as the \LAE must overshine the forward-shock 
emission in order to produce an optical light-curve tracking the X-ray. Such bright optical afterglows will require 
a soft Envelope spectrum with a low-energy slope of $\beta_o^{(E)}= -1/2$, which implies that the X-ray and optical 
flux decay indices should differ by $\alpha_x-\alpha_o=1/4$.

 Therefore, afterglows with decoupled optical and X-ray light-curves require a dual interpretation, with the X-ray 
arising from the \LAE and the optical being produced by the reverse and forward external shocks. The only reason
for which X-rays are more readily attributed to the \LAE and not to the external-shock is our expectation that the
power-per-decade $\nu F_\nu$ of the former model is closer to the X-rays than for the latter.

\acknowledgments{ {\sl Acknowledgments:}
  This work made use of data supplied by the UK Swift Science Data Centre at the University of Leicester:
  (Evans et al 2010) {\sl www.swift.ac.uk/burst\_analyser} 
  and was supported by an award from the LDRD program at the Los Alamos National Laboratory}

\begin{figure}
%\centerline{\psfig{figure=GtDLC.eps,width=18cm,height=14cm}}
\centerline{\includegraphics[width=18cm,height=14cm]{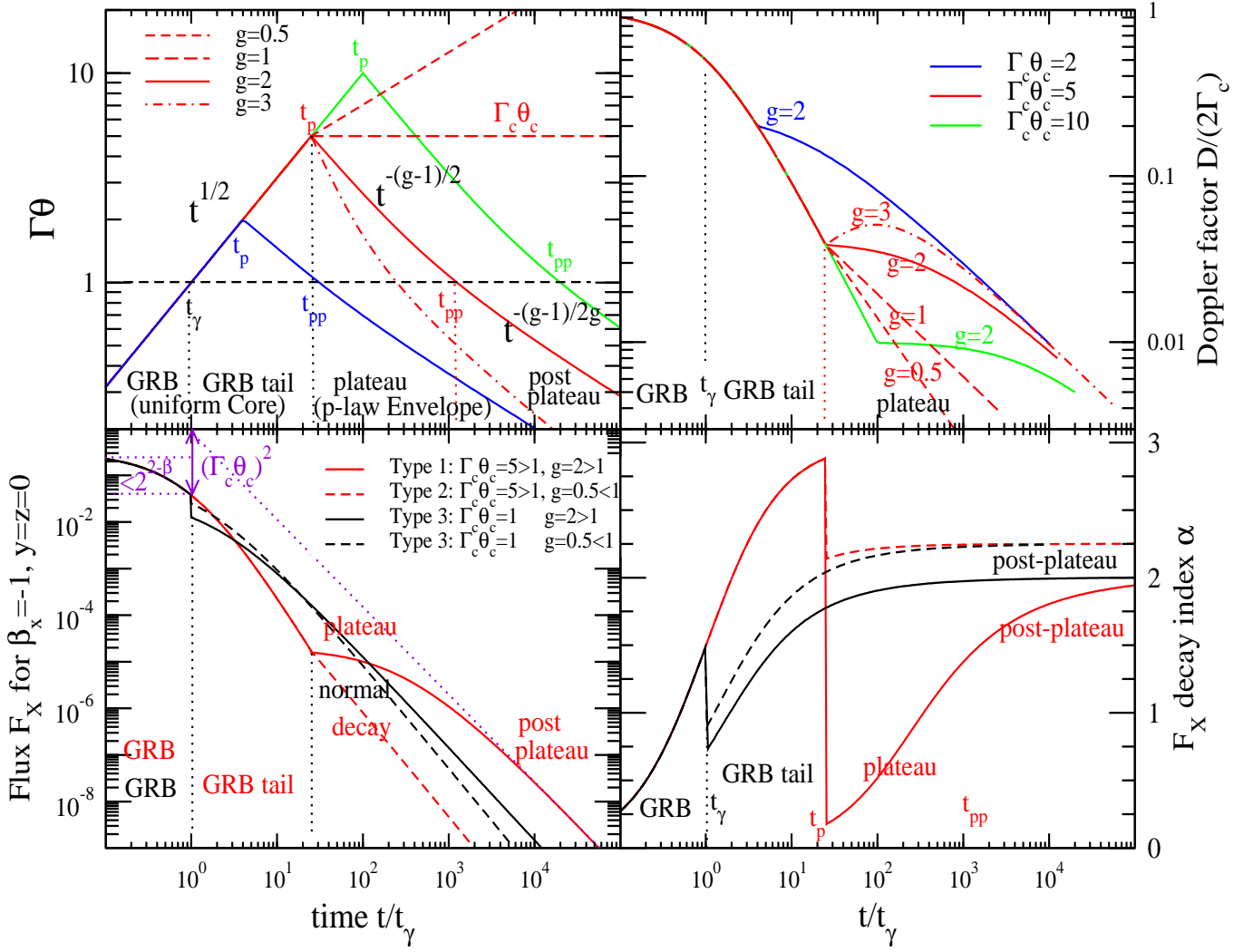}}
\vspace*{5mm}
\figcaption{ \normalsize
 {\bf Upper panels}: 
  Evolution of two quantities of interest pertaining to the relativistic beaming of the \LAE for three values of the Core 
 parameter $\Gamma_c \theta_c$ (which sets the duration of the GRB tail and of the afterglow plateau) and four values of 
 the Envelope Lorentz factor index $\Gamma \sim \theta^{-g}$ (which determines how fast the plateau flux evolves). 
 The indicated timescales are: $\tg$ the end of the GRB pulse, $t_p$ the end of the very fast-decaying 
 GRB tail and beginning of the slow-decaying afterglow plateau, $t_{pp}$ the end of the afterglow plateau and
 beginning of the normal aferglow decay. Expected relations: $t_p/\tg = t_{pp}/t_p = (\Gamma_c\theta_c)^2$  
 (last equality for $g=2$).
  {\bf Left upper panel}: $\Gamma \theta$ evolution. If below unity (as for the GRB prompt and post-plateau), 
 then the emission is beamed toward the observer; if above unity (as for the GRB tail and the afterglow plateau), 
 then emission is beamed toward the observer. For $g \siml 1$, the phase $\Gamma \theta > 1$ persists after $t_p$.
  {\bf Right upper panel}: Doppler factor $D$ evolution.  
 During the GRB phase ($t < \tg$), $D$ falls slowly from $2\Gamma_c$ to $\Gamma_c$, leading to a flat prompt flux.
 During the GRB tail ($\tg < t < t_p$), $D \sim t^{-1}$, leading to a very fast decay of the prompt emission.
 During the afterglow plateau ($t_p < t < t_{pp}$), $D \sim t^{(g-2)/2}$, leading to a flat plateau for $g=2$.
 After the plateau ($t_{pp} < t$), $D \sim t^{-1/2}$, leading to a fast post-plateau decay for a uniform source
 surface emissivity.
  {\bf Lower panels}: 
 X-ray light-curves for the comoving-frame emissivity given in equation (\ref{yz}) with $\beta=-1$ and uniform
 spectral peak-flux $i'_p$ ($y=0$) and peak-energy $\nu'_p$ ($z=0$), i.e. independent of angle $\theta$. 
  {\bf Left lower panel}: The Core parameter $\Gamma_c\theta_c$ and the index $g$ determine the morphology of the 
 counterpart and afterglow, leading to three types: type 1 (three post-GRB phases: tail, plateau, post-plateau),
 type 2 (two post-GRB phases: tail and normal decay), type 3 (one post-GRB phase: post-plateau). 
 An afterglow plateau may be missing either because $\Gamma_c \theta_c = 1$ or because $g<1$.
 For afterglow with plateaus, the extrapolation of the post-plateau decay ($t\gg t_{pp}$) to the burst end ($\tg$)
 should be brighter than the flux at that time by a factor $\Gamma^2_c \theta^2_c$ (\S\ref{funny}).
  {\bf Right lower panel}: Evolution of the power-law decay index (logarithmic derivative) for the fluxes in 
 the left panel. The indices given in equation (\ref{alfa}) are recovered asymptotically, late in the GRB tail, 
 at the beginning of the plateau (if the tail and plateau are well-developed: $\Gamma_c\theta_c \simg 5$), 
 or late during the post-plateau: for GRB tail - $\alpha_{tail} = 2-\beta = 3$, for plateau - $\alpha_p = 
 (1-g/2)(2-\beta) = 3(1-g/2)$ if $g > 1$, for post-plateau - $\alpha_{pp} = 2-\beta/2-1/g = 2.5 - 1/g$ if 
 $g > 1$ and $\alpha_{pp} = \alpha_p = 3(1-g/2)$ if $g < 1$.
}
\end{figure}

\begin{figure*}
%\centerline{\psfig{figure=061121.eps,width=14cm,height=14cm}}
\centerline{\includegraphics[width=14cm,height=14cm]{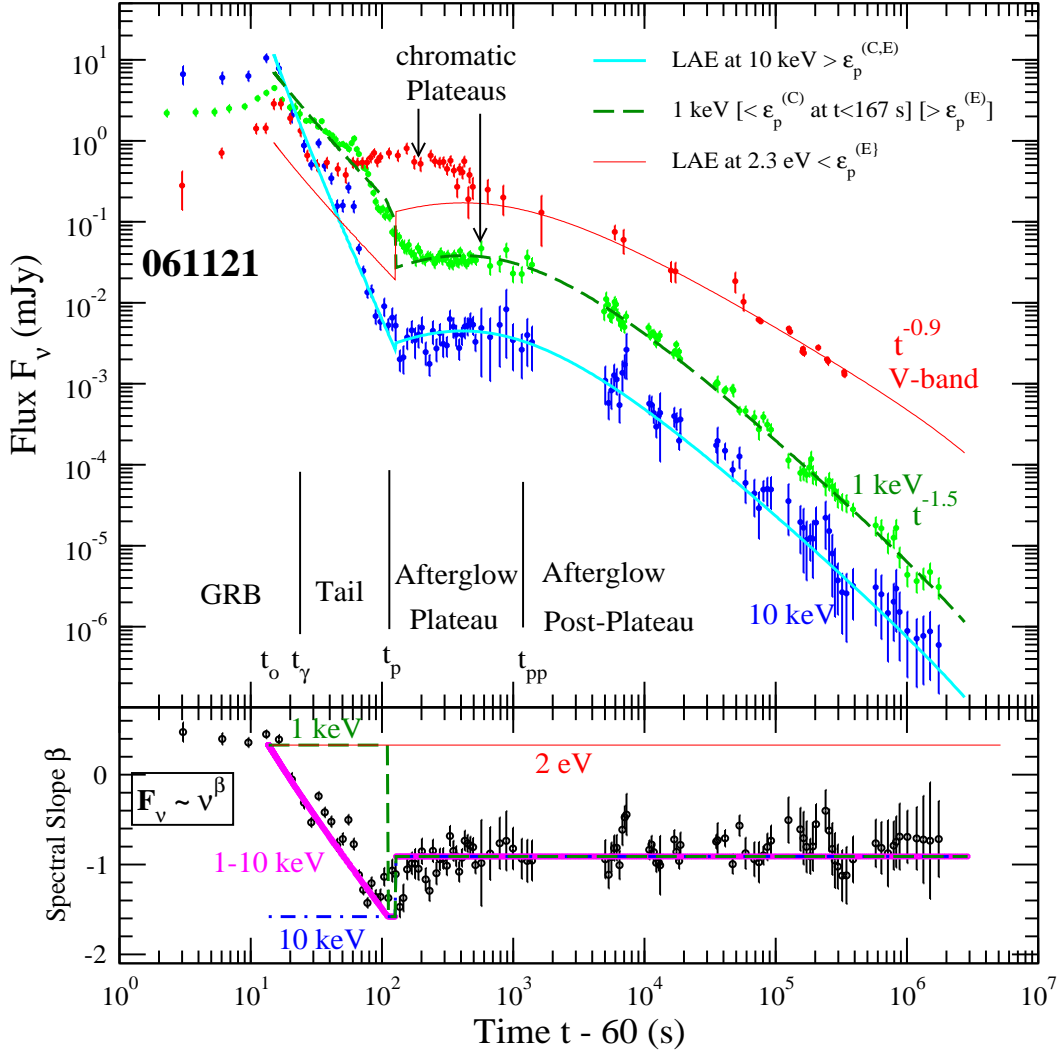}}
\vspace*{5mm}
\figcaption{ \normalsize
 Optical and X-ray prompt (counterpart) and delayed (afterglow) emissions for {\bf GRB 061121}, as measured by 
Swift/XRT and UVOT telescopes (Page et al 2007), plus MDM data for the late optical afterglow, and the 
best-fit obtained with the \LAE model. This is a burst with a GRB tail and an afterglow plateau ({\bf type 1}).
 {\bf Upper panel}: data and model fluxes. 
 For the best-fit, the uniform Core parameter is $\Gamma_c \theta_c = 3.3$ and the last pulse has peak epoch 
$t_o = 75$ s and duration $\tg-t_o = 10$ s.
% The plateau begins at $t_p-t_o = 114$ s and ends at $t_{pp}-t_o = 1150$ s. 
 The Core emission best-fit parameters are: spectral peak-energy $\eps_p^{(C)} = 10$ keV,
peak-flux $F_p^{(C)} = 20$ mJy, both in the observer frame and for an emission boosted by the axis Doppler factor 
$D(\theta=0) = 2\Gamma_c$, spectral slope below peak $\beta_o^{(C)} = 1/3$ (for synchrotron emission from 
uncooled electrons), above peak $\beta_x^{(C)} = -1.6$. 
 The power-law Envelope has a Lorentz factor with an angular dependence $\Gamma \sim \theta^{-2.0}$ and
emission parameters $\eps_p^{(E)} = 0.7 (\theta/\theta_c)^{0.4}$ keV, $F_p^{(E)} = 70 (\theta/\theta_c)^{1.5}$ mJy 
(for the axial Doppler factor above), and spectral slopes $\beta_o^{(E)} = 1/3$ and $\beta_x^{(E)} = -0.93$,
thus the co-moving frame X-ray emissivity satisfies $\inu \sim \theta^{1.9}$.
 The optical and X-ray light-curve plateaus are not simultaneous, which indicates that (at least) the optical plateau
originates from another mechanism.
 {\bf Lower panel}: spectral slopes.
 10 keV is always above the two peak-energies, 1 keV is crossed by the Core's peak-energy just before the plateau 
begins and is above the Envelope's peak-energy at all times, while the optical is always below any peak-energy. 
Thus, the 1-10 keV spectral slope measured at the end of the GRB tail sets the Core's high-energy slope 
$\beta_x^{(C)}$, while that measured during the afterglow sets the Envelope's high-energy slope $\beta_x^{(E)}$. 
 The decrease of the Doppler factor during the GRB tail yields a decrease of the observer-frame spectrum peak-energy 
by a factor $(\Gamma_c\theta_c)^2 + 1 \simeq 12$
and a progressive softening of the 1-10 keV Core spectrum from the low-energy slope $\beta_o^{(C)} = 1/3$ at the
pulse peak (when $\eps_p^{(C)} = 10$ keV) to the low-energy slope $\beta_x^{(C)} = -1.6$ at the end of the GRB tail 
(when $\eps_p^{(C)} = 1$ keV), consistent with the softening of the 1-10 keV spectrum measured by Swift/XRT.
}
\end{figure*}

\begin{figure*}
%\centerline{\psfig{figure=060607.eps,width=14cm,height=14cm}}
\centerline{\includegraphics[width=14cm,height=14cm]{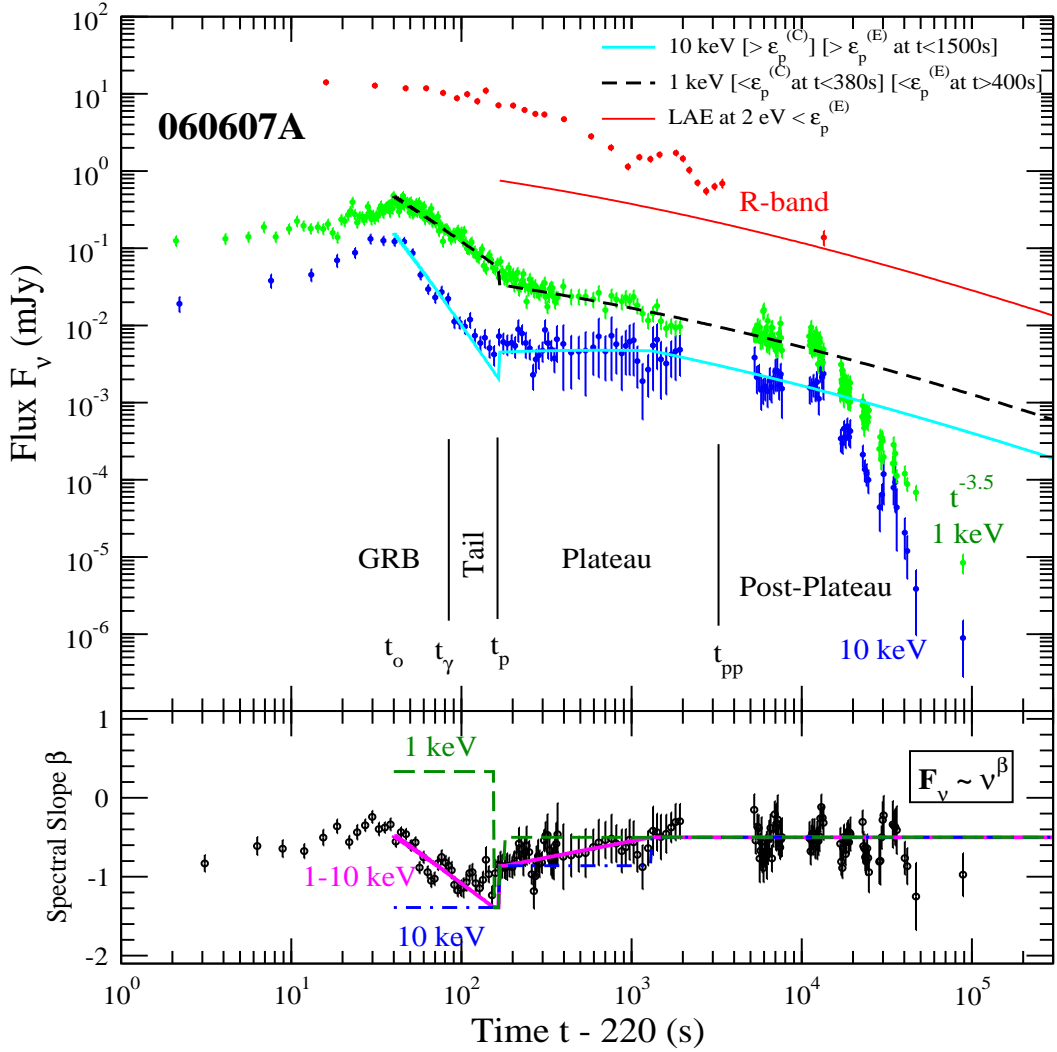}}
\vspace*{5mm}
\figcaption{ \normalsize
 {\bf Upper panel}:
 Swift/XRT counterpart/afterglow and optical data (Nysewander et al 2009) for {\bf GRB 060607A} ({\bf type 1}).
The longevity of its plateau indicates that the power-law index of the Envelope's Lorentz factor distribution
is below the expected "plateau value" $g=2$. The best-fit has $\Gamma \sim \theta^{-1.3}$ and places the plateau 
end earlier (at 3.5 ks) than shown by the X-ray light-curve (fast-decaying post-plateau starts at 10 ks).
 Other parameters for the \LAE model are $\Gamma_c \theta_c = 1.6$, pulse peak epoch $t_o=260$ s, duration 
$\tg-t_o = 48$ s, uniform Core spectral parameters: peak-energy $\eps_p^{(C)} = 3$ keV, peak-flux $F_p^{(C)}= 0.7$ mJy 
for the axial ($\theta=0$) Doppler factor, spectral slope $\beta_x^{(C)} = -1.4$ above the peak-energy.  
 Envelope best-fit parameters are: break-energy $\eps_p^{(E)} = 3 (\theta/\theta_c)^{3.5}$ keV, break-energy 
flux $F_p^{(E)} = 0.7 (\theta/\theta_c)^{0.2}$ mJy, high-energy spectral slope $\beta_x^{(E)} = -0.9$. 
 The Envelope peak spectral parameters are constrained by the hardening seen during the plateau. 
% The spectral softening of the Core's 1-10 keV emission during the GRB tail and the spectral hardening of the 
%Envelope's 1-10 keV emission during the afterglow plateau imply that the Core and Envelope peak spectral parameters 
%are continuous at the Core-Envelope boundary.
 The post-plateau X-ray flux decay  $F_x \sim t^{-3.5}$ is too fast for the Envelope emissivity that fits the plateau 
-- $i'_{1k} \sim \theta^{2.0}$ and $i'_{10k} \sim \theta^{3.2}$ -- and requires a substantial change in the emissivity
angular dependence to $\inu \sim \theta^{-4.7}$ or the emitting surface extends only to an angle $\theta_{max} =
\theta (10\,ks) \siml 6\deg$, so that the sharp flux decay after 10 ks shows how the emission from $\theta_{max}$ 
switches off.
 The optical light-curve cannot be attributed to the \LAE because it is not coupled with the X-ray and lacks a break
at the end of the GRB tail. 
%If the spectral hardening during the X-ray plateau is not imposed on the best-fit, then there are Envelope peak 
%spectral parameters for which the \LAE matches well the optical afterglow measurements, but the optical counterpart 
%data cannot be explained with the Core emission.
 {\bf Lower panel}: 
 During the GRB phase ($t_o-\tg$), the Doppler factor decreases by a factor $\Gamma_c^2\theta_c^2 + 1 = 3.6$, thus
the observer-frame Core peak-energy $\eps_p^{(C)}$ does not sweep the entire 1-10 keV XRT window. Consequently,
to account for the moderate spectral softening seen during the prompt phase requires a hard low-energy spectral 
slope $\beta_o^{(C)} = 1/3$ (uncooled electrons).
 The spectral hardening during the X-ray plateau, from $\beta_x \simeq -1$ to $\beta_x \simeq -1/2$, indicates 
that the peak-energy of the Envelope spectrum increases and that the low-energy spectral slope is $\beta_o^{(E)} = -1/2$
(cooled electrons).
}
\end{figure*}

\begin{figure*}
%\centerline{\psfig{figure=061110.eps,width=14cm,height=14cm}}
\centerline{\includegraphics[width=14cm,height=14cm]{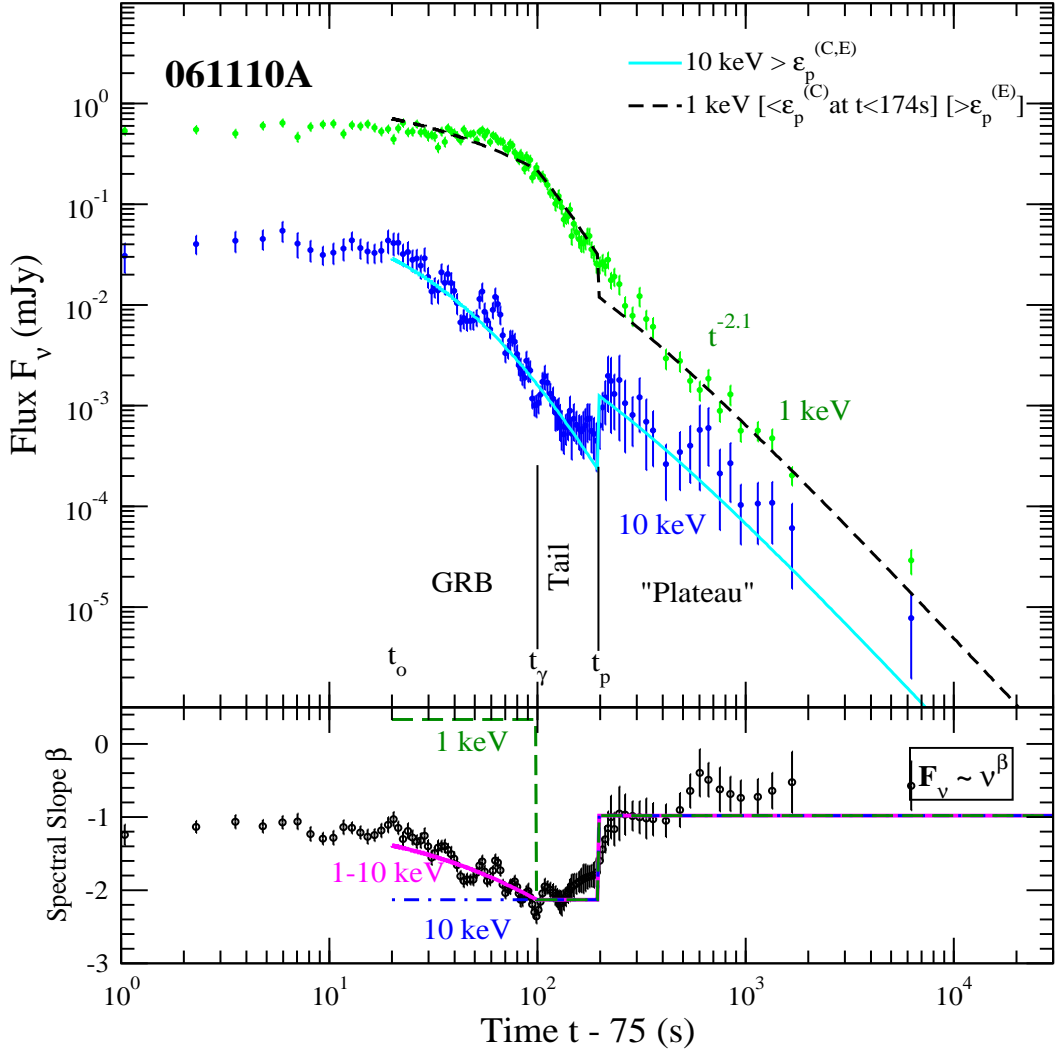}}
\vspace*{5mm}
\figcaption{ \normalsize
 {\bf Upper panel}:
 X-ray counterpart and afterglow of {\bf GRB 061110A}. 
 This burst shows a tail but the afterglow has only a normal decay phase ({\bf type 2}), lacking a plateau, 
which indicates that the Envelope Lorentz factor distribution index $\Gamma \sim \theta^{-g}$ has $g < 1$.
 In this case, the afterglow is in the $\Gamma\theta > 1$ phase (radiation beamed away from the observer), 
however the light-curve displays a normal decay because the Doppler factor decreases fast (Figure 1).
 Core parameters are: $\Gamma_c\theta_c = 1.5$, peak epoch $t_o = 95$ s, duration $\tg-t_o = 80$ s, observer-frame 
peak-energy $\eps_p^{(C)} = 2$ keV, peak-flux $F_p^{(C)} = 0.9$ mJy for $\theta=0$, spectral slope above peak 
$\beta_x^{(C)} = -2.1$.
 Envelope parameters: index $g=0.5$ (constrained only by the afterglow flux decay), peak-energy $\eps_p^{(E)} = 1.7$ keV, 
peak-flux $F_p^{(E)} = 0.3$ mJy, both set to be uniform, and high-energy spectral slope $\beta_x^{(E)} = -1.0$. 
The Envelope peak spectral parameters and their angular dependence are poorly constrained when only X-ray data 
are modeled, as they are in fact degenerate. Instead, fits to light-curves at an energy above the peak $\eps_p^{(E)}$ 
constrain the quantity $F_p/\eps_p^\beta$ and its angular distribution. 
 {\bf Lower panel}: 
 As for GRB 060607A, the observer-frame Core peak-energy $\eps_p^{(C)}$ decreases by a factor $\Gamma_c^2\theta_c^2+1=3.2$ 
during the burst, it does not cross the entire 1-10 keV XRT window, and accounting for the moderate spectral softening 
during the prompt phase requires a harder low-energy spectral slope $\beta_o^{(C)} = 1/3$ (uncooled electrons), as the 
resulting softening for the softer slope $\beta_o^{(C)} = -1/2$ (cooled electrons) would be too small compared with what 
is measured.
}
\end{figure*}

\begin{figure*}
%\centerline{\psfig{figure=061007.eps,width=14cm,height=14cm}}
\centerline{\includegraphics[width=14cm,height=14cm]{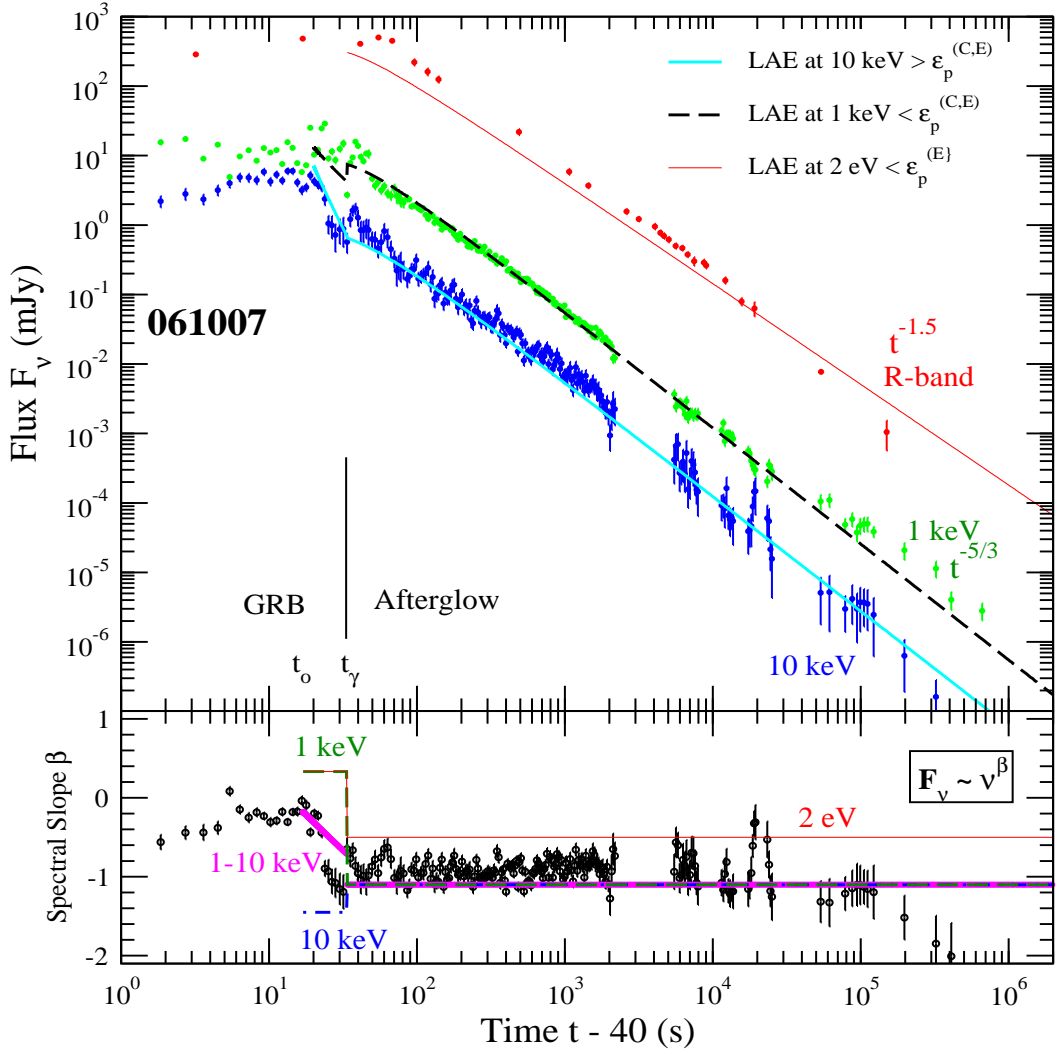}}
\vspace*{5mm}
\figcaption{ \normalsize
 {\bf Upper panel}:
 X-ray counterpart/afterglow measured by Swift/XRT and optical afterglow (ROTSE -- Yost et al 2007 and FST -- 
Mundell et al 2007) of {\bf GRB 061007}.
 This burst does not exhibit a tail and the afterglow does not have a plateau ({\bf type 3}), which indicate that 
$\Gamma_c\theta_c=1$. 
 Best-fit model parameters for the last dominant pulse are: peak epoch $t_o = 57$ s, duration $\tg-t_o = 17$ s,
 Core parameters: observer-frame peak-energy $\eps_p^{(C)} = 5$ keV, peak-flux $F_p^{(C)}= 18$ mJy for $\theta=0$, 
spectral slope above peak $\beta_x^{(C)} = -1.45$.
 X-ray and optical flux decay indices are insufficient to determine the index $g$ of the Envelope Lorentz factor 
angular distribution $\Gamma \sim \theta^{-g}$ and the two indices for the angular distribution of the spectral peak 
characteristics, thus we have assumed $g=2$.
 Other Envelope parameters are: break-energy $\eps_p^{(E)} = 1.0 (\theta/\theta_c)^{0.1}$ keV, break-energy flux 
$F_p^{(E)} = 180 (\theta/\theta_c)^{1.0}$ mJy, high-energy spectral slope $\beta_x^{(E)} = -1.10$. 
 {\bf Lower panel}: 
 Lacking a GRB tail, the Doppler factor and the observer-frame Core peak-energy decrease by only a factor 2 during
the burst phase ($t_o-\tg$). Consequently, a hard low-energy Core spectrum of slope $\beta_o^{(C)} = 1/3$ is needed
to acquire the largest possible softening during the burst, but even that seems less than what is measured. 
 The high brightness of the optical counterpart requires an Envelope low-energy slope $\beta_o^{(E)} = -1/2$,
characteristic of synchrotron emission from cooled electrons. A second constraint on the Envelope low-energy slope
comes from the measured decay of the optical flux.
}
\end{figure*}

\end{document}